\documentclass[sigconf]{acmart}

\usepackage{amsmath}
\usepackage{amsfonts}
\usepackage{microtype}
\usepackage{multirow}
\usepackage{graphicx}
\usepackage{subfigure}
\usepackage{booktabs} 

\AtBeginDocument{%
  \providecommand\BibTeX{{%
    \normalfont B\kern-0.5em{\scshape i\kern-0.25em b}\kern-0.8em\TeX}}}

\begin{document}

\title{Multi-Density Attention Network for Loop Filtering in Video Compression}


\author{Zhao Wang, Changyue Ma, Yan Ye}
\affiliation{%
  \country{Alibaba Group}
 }
\email{{baixiu.wz, changyue.mcy, yan.ye}@alibaba-inc.com}



\begin{abstract}
Video compression is a basic requirement for consumer and professional video applications alike. Video coding standards such as H.264/AVC and H.265/HEVC are widely deployed in the market to enable efficient use of bandwidth and storage for many video applications. To reduce the coding artifacts and improve the compression efficiency, neural network based loop filtering of the reconstructed video has been developed in the literature. However, loop filtering is a challenging task due to the variation in video content and sampling densities. In this paper, we propose a on-line scaling based multi-density attention network for loop filtering in video compression. The core of our approach lies in several aspects: (a) parallel multi-resolution convolution streams for extracting multi-density features, (b) single attention branch to learn the sample correlations and generate mask maps, (c) a channel-mutual attention procedure to fuse the data from multiple branches, (d) on-line scaling technique to further optimize the output results of network according to the actual signal.
The proposed multi-density attention network learns rich features from multiple sampling densities and performs robustly on video content of different resolutions. Moreover, the on-line scaling process enhances the signal adaptability of the off-line pre-trained model. 
Experimental results show that 10.18\% bit-rate reduction at the same video quality can be achieved over the latest Versatile Video Coding (VVC) standard. The objective performance of the proposed algorithm outperforms the state-of-the-art methods and the subjective quality improvement is obvious in terms of detail preservation and artifact alleviation.

\end{abstract}



\maketitle

\section{Introduction}
\label{Introduction}
Digital video is one of the most important multimedia carrier for
humans to acquire information and perceive the world. The video industry has continued to develop rapidly in the forms of many popular services, e.g., on-demand streaming, video conferencing, social networking, e-commerce, remote education, and so on. Due to huge data volume, video compression/coding plays a crucial role to reducing the storage space and transmission bandwidth. Video coding standards, such as the H.264/Advanced Video Coding (AVC) standard \cite{H.264} and the successive High Efficiency Video Coding (HEVC) standard \cite{HEVC}, are widely deployed on many devices and used in many video applications. Ever increasing pursuit of high resolution quality and new media applications present higher requirements and new challenges for video compression. Over the last years, the Joint Video Exploration Team (JVET), a joint committee of ISO/IEC Moving Picture Experts Group (MPEG) and ITU-T Video Coding Experts Group (VCEG), has been working on the state-of-the-art video compression standard Versatile Video Coding (VVC) \cite{VVC}. The first version of the VVC standard was released in July 2020, and achieves up to 40\% bit-rate reduction while maintaining the same quality compared with HEVC.

To reduce the coding artifacts in compressed video, such as blocking artifacts, ringing artifacts, banding, color shift and so on, loop filtering on the reconstructed frames before output has been adopted in video coding standards. For the raw reconstructed frames, loop filtering is applied to reduce the compression distortion and remove the artifacts. Therefore, reconstructed frames in higher quality are produced and better reference can be provided for coding of the subsequent frames. Specifically, deblocking filter (DBF) \cite{deblocking} is widely used to smooth the samples around the block boundaries and remove or reduce the blocking artifacts introduced by the block-based coding structure. The Sample Adaptive Offset (SAO) filter \cite{SAO} was first adopted in HEVC to reduce sample distortion by classifying reconstructed samples into different categories, obtaining an offset for each category, and then adding the offset to each sample of the category. Inspired by the Wiener filter, Adaptive Loop Filtering (ALF) \cite{ALF} introduced in VVC minimizes the mean square error between original samples and reconstructed samples by adapting filter coefficients to the video signal. Besides alleviating subjective artifacts, SAO and ALF can achieve about 1.2\% and 4.0\% bit-rate savings respectively. 

Recently, the use of deep neural networks for image restoration/enhancement and image deblurring/denoising tasks has advanced significantly. This motivates people to introduce neural network into video compression as loop filtering to reduce the compression distortion through learning the mapping between the reconstructed frame and the original frame. Specifically, Park and Kim first proposed a convolutional neural network (CNN) in loop filtering \cite{park2016cnn} based on the extension of CNN for super resolution \cite{dong2014learning}. Compared with a single-path CNN, a CNN with variable filter sizes was proposed to help extract features in different scales \cite{dai2017convolutional}. A multi-channel long-short-term dependency residual network was proposed by introducing an update cell to adaptively store and select the long-term and short-term dependency information \cite{meng2018new}. Besides the sample values in reconstructed frames, partitioning information during compression was included as additional input in \cite{kang2017multi} and the spatial-temporal neighbouring samples were utilized in \cite{jia2017spatial}. Wang et al. \cite{wang2019attention} further adopted the attention based dual-scale CNN in loop filtering and achieved promising performance. To learn an enriched set of features, multi-branch network was proposed for image restoration and enhancement \cite{zamir2020learning}. Moreover, on-line training based loop filtering was also investigated \cite{lin2020towards}.

However, there are some issues in the existing CNN based methods. Firstly, the limitation lies in the trade-off between spatially-precise representations and larger receptive field. Generally, video compression requires precise signal-level information. Down-sampling is a commonly used step in other computer vision and visual analysis tasks but it could introduce unwanted loss of signal details in video compression. On the other hand, full-resolution convolution layer is less effective in utilizing contextual information due to their limited receptive field. Secondly, it is observed that a model trained offline on a certain resolution tends to perform worse when the test video is in a different resolution. Considering the video resolution represents the sampling density of the original signal, video in various resolutions usually has different degrees of sample-correlation within a given neighborhood. Therefore, the ability to adapt to different resolutions should be an important feature of the model for loop filtering. Thirdly, how to fuse the features extracted from multiple branches effectively is a challenging problem. Concatenating or element-wise sum/product is commonly used but could incur performance loss. Finally, the off-line pre-trained model can not be sufficient for any input signal because of the data diversity. If the original signal can be utilized during inference, the performance of loop filtering can be improved.  

To address the above issues, we propose the multi-density attention network for loop filtering based on the core block in a multi-density single-attention structure. In each block, multi-density branches are designed to extract the features in various scales and an attention branch is introduced to learn the sample correlations. The data from multiple branches are further fused based on the channel-mutual attention procedure to generate the output. Furthermore, we propose the on-line scaling method to enhance the performance of off-line trained model when inferred in the video codec.
The main contribution of this work are summarized:  
\begin{itemize}
\setlength{\itemsep}{0pt}
\item A novel multi-density feature extraction model that obtains complementary set of features across multiple spatial scales, simultaneously maintaining the original high-resolution features to preserve precise details and utilizing the larger receptive field. 
\item The multi-density single-attention structure is proposed, where the single attention branch can learn the spatial sample correlations and generate multiple mask maps to refine the feature maps. 
\item A new approach to fuse the data from multiple branches, which dynamically selects the useful set of features from each branch using a channel-mutual attention mechanism.
\item A on-line scaling technique to optimize the output results of the neural network and further reduce the distortion between the filtered frame and the original frame.
\end{itemize}

\section{Related Work}

\textbf{Loop filtering.} Many literature investigated the incorporation of neural network into video compression \cite{jia2019content, pan2020efficient}, the utilization of signal and compression information \cite{kang2017multi, jia2017spatial} and the multi-frame filtering structure \cite{ lee2018convolution, li2019deep}. In JVET, an ad-hoc group on neural network based video coding was established and some methods regarding loop filtering were proposed \cite{JVET-T0079,JVET-T0088,JVET-T0094,JVET-U0054}. The model architecture of most of them is similar to the modified ResNet \cite{he2016deep}, where several Resblocks are inserted between the input and output to learn the mapping between reconstructed frame and the original frame. Another approach first applies the down-sampling to the input, then go through the Resblocks and up-sampled before output \cite{JVET-T0088}. However, the single-path model is hard to utilize spatially-precise representations and large receptive field simultaneous. With respect to the inference of neural network for video filtering, frame-level on/off control was investigated in \cite{ding2019switchable} and an efficient finetuning methodology was proposed to adapt the neural network to the specific content \cite{lam2020efficient}.     

\textbf{Attention mechanism.} Generally, attention can be viewed as a guidance to bias the allocation of available processing resources towards the most informative components of features. It’s usually combined with a gating function (e.g., sigmoid) to rescale the feature maps. Wang et al. \cite{wang2017residual} proposed residual attention network for image classification with a trunk-and-mask attention mechanism. Hu et al. \cite{hu2018squeeze} proposed squeeze-and-excitation (SE) block to model channel-wise relationship to obtain significant performance improvement for image classification. In all, these works mainly aim to guide the network's attention towards the regions of interest. However, few work has been proposed to investigate the attention mechanism to learn the spatial correlation of various sampling densities for improved video compression efficiency. 

\textbf{Enhancement/restoration.} Image enhancement and restoration usually focus on low-quality images, such as low-light image, noisy image, blurry image and so on. Based on the residual learning and batch normalization, Zhang et al. \cite{zhang2017beyond} proposed DnCNN model to handle Gaussian denoising with unknown noise level. By incorporating the non-local block \cite{wang2018non}, Zhang et al. \cite{zhang2019residual} proposed the residual non-local attention network for image restoration to capture the long-range dependencies. The spatio-temporal filter adaptive network was developed for video deblurring \cite{zhou2019spatio} and optical flow estimation was jointly trained with processing component for video enhancement \cite{xue2019video}. Recently, generative adversarial network (GAN) also attracted considerable attention due to its favorable performance especially for extremely low bit-rate compression \cite{deng2018aesthetic,ma2020cvegan}.

\section{Multi-Density Attention Learning}

We first give a general definition of multi-density attention learning and then present our implemented instantiation of it.

\subsection{Formulation}

The CNN model is designed to learn the mapping between the input and the output by convolution layer. The convolution operation on the representations (feature maps) in the hidden layers can be defined as: 
\begin{equation} \label{eq:1}
y_i = \sum_{j\epsilon \Omega_i} w_j \cdot x_j,
\end{equation}
where $i$ is the index of an output position and $j$ is the index that enumerates local positions $\Omega_i$ centering at $i$. $x$ and $y$ denote the input and output representations respectively and $w$ are the parameters to be learned.

Considering the parameters trained on a given dataset, they converge to optimize the mapping between input and output in terms of the overall data. When the pre-trained model is used in inference, it may adapt poorly to the specific input signal. In other words, a model trained on a specific content could perform worse on the data with different types of content. Even for the same content, a model trained on a specific resolution may not be the most effective for the same input but in other resolutions, because various resolutions represent different sampling densities. To enhance the model adaptability regarding input signal,the sample-correlation should be taken into consideration in the convolution operation as:
\begin{equation} \label{eq:2}
y_i =  \sum_{j\epsilon \Omega_i} f(x_i, x_j) \cdot w_j \cdot x_j.
\end{equation}
Here, a pairwise function $f(x_i, x_j)$ represents the correlation between $x_i$ and $x_j$. 

\begin{figure}[t]
\vskip 0.2in
\begin{center}
\centerline{\includegraphics[width=7.5cm]{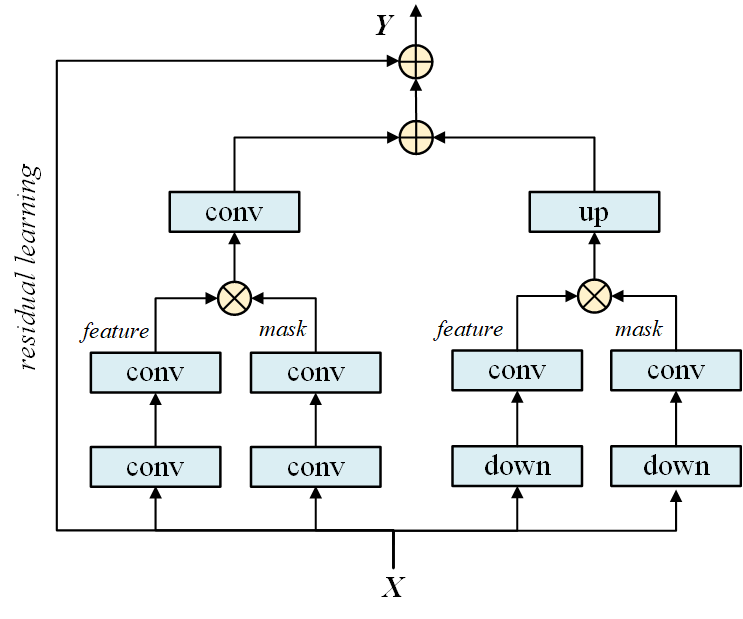}}
\caption{Illustration of the multi-density attention learning. \emph{conv}, \emph{up} and \emph{down} represent the convolution layer, down-sample and up-sample process, respectively.}
\label{fig1}
\end{center}
\vskip -0.2in
\end{figure}

However, the pairwise correlation function demands huge number of parameters, such as a full-connection layer. To address this issue, we modify the mutual correlation $f(x_i, x_j)$ to the attention factor $F(x_i)$, which represents the general influence of local neighborhood on $x_i$. In this manner, the sample-correlation convolution operation can be expressed as:
\begin{equation} \label{eq:3}
y_i =  F(x_i) \cdot  \sum_{j\epsilon \Omega_i} (w_j \cdot x_j).
\end{equation}
Let $G(x_i)$ denote the convolution response in \eqref{eq:1}, the above equation can be rewritten as:
\begin{equation} \label{eq:4}
y_i =  F(x_i) \cdot G(x_i).
\end{equation}

Due to the signal diversity, the potential ability to adapt will be limited if the sample correlation is learned within a fixed local neighborhood, e.g. $3 \times 3$ kernel size. Therefore, we propose to learn the features and correlations in multi-density neighborhoods, which can be expressed as:
\begin{equation} \label{eq:5}
y_i =  \frac{1}{N(x)}  \sum_{\lambda \epsilon \Theta} F_{\lambda}(x_i) \cdot G_{\lambda}(x_i),
\end{equation}
where $\lambda$ $\epsilon$ $\Theta$ denotes the density levels. After the multi-density responses are collected, they are normalized by a factor $N(x)$. The multi-density learning can be realized by adopting different kernel sizes or various resolutions.

To enhance the learning ability in deep neural network, the residual learning by an additional shortcut connection \cite{he2016deep} is incorporated. Thus we modify the multi-density attention operation as: 
\begin{equation} \label{eq:6}
y_i = x_i + \frac{1}{N(x)}  \sum_{\lambda \epsilon \Theta} F_{\lambda}(x_i) \cdot G_{\lambda}(x_i),
\end{equation}

\begin{figure*}[t]
\vskip 0.2in
\begin{center}
\centerline{\includegraphics[width=17cm]{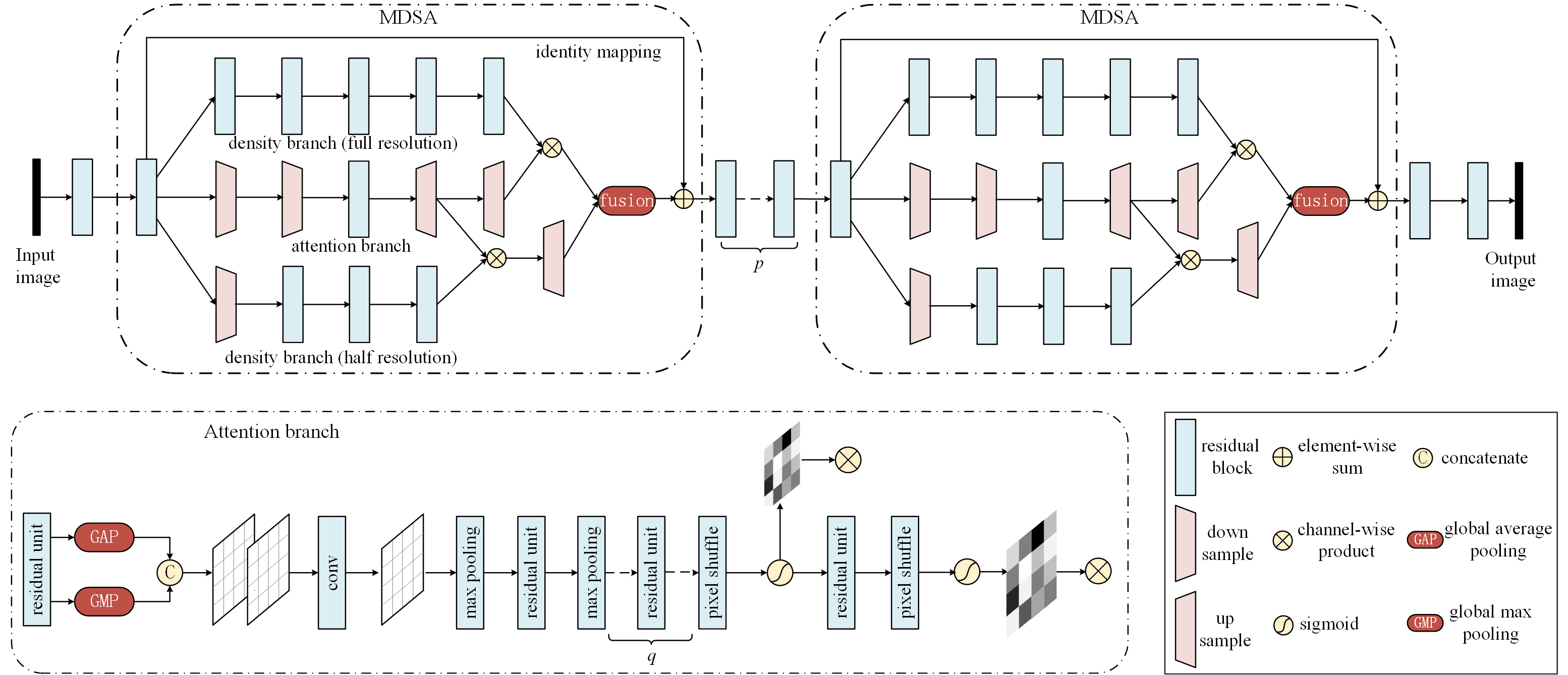}}
\caption{Illustration of the proposed MDAN. We use two hyper-parameters here: \emph{p} and \emph{q}, where \emph{p} denotes the number of residual blocks between two MDSA blocks and \emph{q} represents the number of residual blocks between the last pooling layer and the first pixel-shuffle layer in the attention branch. In our experiments, we set  $\{\emph{p} = 2, \emph{q} = 1$\}.}
\label{fig2}
\end{center}
\vskip -0.2in
\end{figure*}

The flowchart of multi-density attention operation is illustrated in Figure \ref{fig1}, where only two densities are taken as an example. The left branches are based at the full-resolution scale while the right branches are processed at the half-resolution scale. In each density scale, the feature maps and the attention masks are fused by element-wise product. Then, the responses from different densities are fused by element-wise sum, the identity mapping is added, and finally the output is generated. 

\subsection{Multi-Density Attention Network}

Architecture of the proposed multi-density attention network (MDAN) is illustrated in Figure \ref{fig2}. For the input image, the representations after one residual block go through several multi-density single-attention (MDSA) blocks to obtain the output, and the number of MDSA block is set as a hyper-parameter. The core design of the MDSA block contains three elements: (a) multi-density branches for extracting rich features in various receptive fields and spatial resolutions, (b) a single attention branch to generate multiple mask maps to refine the features with spatial attention mechanism being introduced to reduce the complexity, (c) a novel channel-mutual attention based fusion module for feature aggregation. 

\textbf{Multi density branches.} 
In order to extract the representations of the input image, existing CNNs typically employ the following architecture design: (a) the receptive field of neurons is fixed in each layer and the feature maps are kept at the same full resolution as the input size, (b) the spatial size of feature maps is gradually reduced (and later progressively recovered) to utilize larger neighborhood and generate a semantically strong representation. When the above architecture is adopted to perform loop filtering, such design cannot benefit from larger receptive field and precise representations simultaneously, because the first method is limited by the receptive field and the second method introduces unwanted loss of signal by the down-sampling process. Larger kernel size (e.g. $5 \times 5$, $7 \times 7$) such as the Inception network \cite{szegedy2016rethinking} has the advantage of larger receptive field and more precise representation simultaneously. However, the number of parameters of $5 \times 5$ kernel size increases significantly compared to $3 \times 3$ adopted in our implementation, and the $7 \times 7$ kernel size has even higher complexity. 

In the proposed MDSA block, we adopt multiple branches in different spatial resolutions to benefit from larger receptive field and more precise representations. Specifically, the full-density branch is required to preserve the details of signal. The half-density branch is introduced to learn the sample-correlations in larger receptive field. Moreover, quarter-density or even sparser density branches can be added to further enhance the learning ability. In our implementation, only full-density and half-density branches are used to balance the performance and the complexity. Though the kernel size of $3 \times 3$ is still used in half-density branch, the perception covers a larger neighbourhood of sample area in the original image.  

As shown in Figure \ref{fig2}, the full-density branch consists of several residual blocks. In each residual block, it contains two convolution layers and the batch normalization layers are removed to avoid losing range flexibility when normalizing the features, as presented in previous works \cite{nah2017deep, lim2017enhanced}. With respect to the half-density branch, a down-sampling layer is firstly applied to obtain the half-resolution representations, followed by several successive residual blocks. Then, the mask map generated in the attention branch is applied to refine the half-density features by spatial-wise product. Finally, an up-sampling layer is applied to resize the feature map back into the original resolution. The down-sampling layer is realized by convolution layer with stride of two and the up-sampling layer is realized using the pixel-shuffle procedure \cite{shi2016real}.

\begin{figure*}[t]
\vskip 0.2in
\begin{center}
\centerline{\includegraphics[width=17cm]{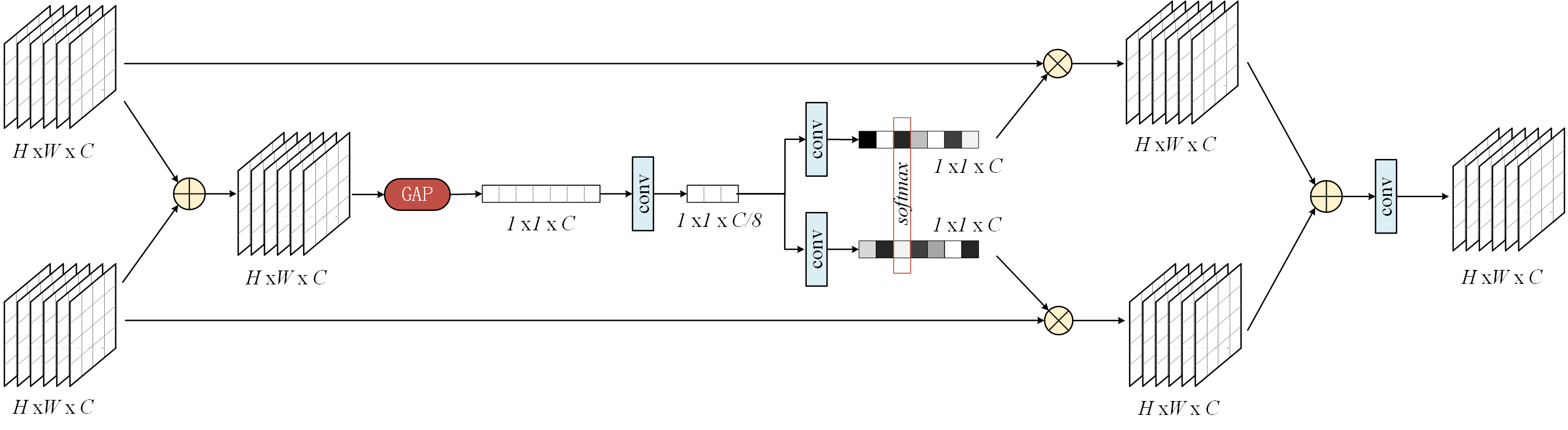}}
\caption{Illustration of the channel-attention based fusion procedure in the proposed MDSA block.}
\label{fig3}
\end{center}
\vskip -0.2in
\end{figure*}

\textbf{Single attention branch.}  
According to the formulation \eqref{eq:5}, we can select to construct one attention branch for each density branch to learn the mask map. Nevertheless, this approach demands additional parameters and having too many branches also increases the model complexity, training time and memory footprint. To address this issue, we propose the multi-density single-attention architecture, which adopts bottom-up top-down structure in the attention branch to generate mask maps in multiple scales to be applied into each corresponding feature map. Moreover, the attention branch aims at learning the sample correlations in various spatial densities and softly weight the density branch features. Therefore, spatial attention mechanism is introduced that focus on the inter-spatial mask, to further reduce the complexity of the attention branch. 

Following previous attention mechanism \cite{wang2017residual}, our attention branch contains fast feed-forward sweep and top-down feedback steps. The former operation quickly collects global information of the whole image, and the latter operation combines global information with original feature maps. Detailed implementation of attention branch is illustrated in Figure \ref{fig2}. For the input $f$ $\epsilon$ $\mathbb{R}^{H \times W \times C}$, global average pooling (GAP) and global max pooling (GMP) are firstly applied independently along the dimension of channels and the outputs are concatenated to form a map of $\mathbb{R}^{H \times W \times 2}$. Then, the map is passed through a convolution to obtain $f^{'}$ $\epsilon$ $\mathbb{R}^{H \times W \times 1}$.
Successively, spatially max pooling and residual block are alternately performed several times to increase the receptive field rapidly. After reaching the lowest resolution, the global information is then expanded by the pixel-shuffle layers. When the map reaches the same size with the half-density branch, a sigmoid operation normalizes the map and generate a mask map $m$ $\epsilon$ $\mathbb{R}^{H/2 \times W/2 \times 1}$ within the range of $[0, 1]$. One one hand, the mask map $m$ is fed into the feature map from the half-density branch to obtain the weighted feature map by spatial-wise product. On the other hand, it will be further expanded by the combination of residual block and sigmoid to produce the full-resolution mask map $\hat{m}$ $\epsilon$ $\mathbb{R}^{H \times W \times 1}$, which will be fused with the output feature map of full-density branch.  

The number of residual blocks between the last down-sampling and the first up-sampling in the attention branch is set as hyper-parameter $q$. In our implementation, $q$ is set to 1. With respect to the density branches, the numbers of residual blocks are set to $\{q+2, q+4\}$ for the half-density branch and full-density branch respectively, namely $\{3, 5\}$ in our experiments.

\textbf{Fusion module.}
One fundamental property of the proposed multi-density attention network lies in the multi-stream fusion, which aims to select proper features and suppress the others. The most commonly used approaches for feature fusion include simple concatenation or summation. However, these methods limit the potential of adaptive feature aggregation and selection. Inspired by the selective kernel network and channel attention\cite{li2019selective, zamir2020learning, wang2017residual}, we introduce a nonlinear procedure for fusing features coming from multiple branches based on the channel-attention mechanism.

The proposed fusion module is shown in Figure \ref{fig3}. The core architecture includes the fast generation of channel masks and the selection of features using mask maps. Specifically, the two inputs from two density branches are first combined by element-wise sum and go through global average pooling across the spatial dimension to obtain channel-wise statistics $s$ $\epsilon$ $\mathbb{R}^{1 \times 1 \times C}$. Next, we apply a channel-downscaling convolution layer to generate a compact feature representation $z$ $\epsilon$ $\mathbb{R}^{1 \times 1 \times r}$, where $r$ is a hyper-parameter and set to $C/8$ in our implementation. Successively, the feature vector $z$ passes through parallel channel-upscaling convolution layer (one for each input) and provides us with two feature descriptors. After softmax function, the channel-wise mask vectors $s_1$ and $s_2$ with size of $1 \times 1 \times C$ are obtained. Applying the mask vectors to the input feature maps respectively, the results are summed and go through one more $1 \times 1$ convolution layer and finally output the softly masked feature map. 

\begin{table*}[tbp]
\caption{The performance of the proposed on-line scaling based multi-density attention network for loop filtering.}
\label{table1}
\small
\begin{tabular}{ccccccccccccc}
\toprule
\multirow{2}{*}{\textbf{Sequence}} & \multicolumn{4}{c}{\textbf{BD-rate (MDAN)}} & \multicolumn{4}{c}{\textbf{BD-rate (MDAN+scaling)}} & \multicolumn{4}{c}{\textbf{BD-PSNR (MDAN+scaling)}}  \\
 & \textbf{Y} & \textbf{U} & \textbf{V} & \textbf{YUV} & \textbf{Y} & \textbf{U} & \textbf{V} & \textbf{YUV} & \textbf{Y} & \textbf{U} & \textbf{V} & \textbf{YUV}  \\ 
\midrule
  \emph{Tango2}          & -7.16\% & -18.11\% & -14.23\% & -10.16\% & -7.78\% & -19.89\% & -16.42\% & -11.24\% & +0.207 & +0.587 & +0.540 & +0.326  \\
  \emph{FoodMarket4}     & -7.53\% & -10.63\% & -10.10\% & -8.48\%  & -8.14\% & -11.76\% & -12.37\% & -9.45\%  & +0.276 & +0.477 & +0.474 & +0.343  \\
  \emph{Campfire}        & -8.13\% & -7.41\%  & -14.75\% & -9.11\%  & -8.62\% & -8.52\%  & -16.41\% & -9.90\%  & +0.248 & +0.460 & +0.567 & +0.337  \\
  \emph{CatRobot}        & -6.81\% & -15.78\% & -15.50\% & -9.75\%  & -7.06\% & -17.43\% & -17.92\% & -10.60\%  & +0.231 & +0.444 & +0.561 & +0.322  \\
  \emph{DaylightRoad2}   & -7.64\% & -20.17\% & -15.40\% & -11.02\% & -8.03\% & -21.36\% & -17.53\% & -11.84\% & +0.202 & +0.576 & +0.457 & +0.307  \\
  \emph{ParkRunning3}    & -2.97\% & -5.05\%  & -8.80\%  & -4.29\%  & -3.69\% & -8.42\%  & -10.82\% & -5.67\%  & +0.137 & +0.314 & +0.329 & +0.199  \\
  \emph{MarketPlace}     & -3.96\% & -17.86\% & -18.37\% & -8.68\%  & -4.66\% & -18.27\% & -19.76\% & -9.45\%  & +0.164 & +0.519 & +0.462 & +0.273  \\
  \emph{RitualDance}     & -7.41\% & -16.54\% & -18.11\% & -10.72\% & -7.94\% & -18.46\% & -20.12\% & -11.72\% & +0.410 & +0.699 & +0.769 & +0.518  \\
  \emph{Cactus}          & -6.39\% & -10.03\% & -12.20\% & -7.97\%  & -6.65\% & -12.15\% & -14.14\% & -8.82\%  & +0.233 & +0.359 & +0.470 & +0.294  \\
  \emph{BasketballDrive} & -6.87\% & -12.17\% & -15.22\% & -9.15\%  & -7.07\% & -15.61\% & -17.88\% & -10.30\%  & +0.225 & +0.475 & +0.600 & +0.329  \\
  \emph{BQTerrace}       & -5.70\% & -16.62\% & -11.45\% & -8.48\%  & -6.09\% & -17.54\% & -12.49\% & -9.07\%  & +0.168 & +0.522 & +0.380 & +0.262  \\
  \emph{BasketballDrill} & -6.05\% & -15.41\% & -17.21\% & -9.47\%  & -6.45\% & -17.13\% & -19.31\% & -10.37\%  & +0.302 & +0.653 & +0.793 & +0.442  \\
  \emph{BQMall}          & -5.16\% & -22.32\% & -19.18\% & -10.36\% & -5.69\% & -23.87\% & -20.55\% & -11.20\% & +0.265 & +0.803 & +0.747 & +0.435  \\
  \emph{PartyScene}      & -6.07\% & -17.90\% & -11.67\% & -8.98\%  & -6.60\% & -19.52\% & -13.83\% & -9.96\%  & +0.305 & +0.644 & +0.480 & +0.391  \\
  \emph{RaceHorses}      & -2.93\% & -18.70\% & -18.72\% & -8.19\%  & -3.63\% & -20.64\% & -20.16\% & -9.22\%  & +0.152 & +0.603 & +0.628 & +0.307  \\
  \emph{BasketballPass}  & -6.33\% & -15.22\% & -15.40\% & -9.32\%  & -6.86\% & -17.26\% & -17.54\% & -10.37\% & +0.355 & +0.716 & +0.746 & +0.480  \\
  \emph{BQSquare}        & -9.01\% & -11.38\% & -26.23\% & -12.28\% & -9.35\% & -12.74\% & -26.49\% & -12.77\% & +0.487 & +0.529 & +1.002 & +0.580  \\
  \emph{BlowingBubbles}  & -4.76\% & -19.16\% & -11.80\% & -8.33\%  & -5.27\% & -20.86\% & -16.85\% & -9.80\%  & +0.221 & +0.727 & +0.497 & +0.351  \\
  \emph{RaceHorses}      & -4.70\% & -24.05\% & -22.20\% & -10.84\% & -5.22\% & -24.85\% & -24.36\% & -11.68\% & +0.250 & +0.864 & +0.856 & +0.453  \\
\midrule
\textbf{Average} & \textbf{-6.08\%} & \textbf{-15.50\%} & \textbf{-15.61\%} & \textbf{-9.24\%} & \textbf{-6.57\%} & \textbf{-17.17\%} & \textbf{-17.63\%} & \textbf{-10.18\%} & \textbf{+0.255} & \textbf{+0.577} & \textbf{+0.598} & \textbf{+0.366}  \\ 
\bottomrule
\end{tabular}
\end{table*}

Due to only $1 \times 1$ convolution is used in our fusion module, the required number of parameters is relatively small compared to the common convolution layer in density branches. For example, if $C = 64$ in the implementation, it requires 5632 parameters in total during the fusion procedure. It is also observed that spatial and inter-channel correlations are exploited in the attention branch and fusion procedure, respectively. The separate spatial- and channel- attention design makes the multi-density attention learning robust against the resolution and content diversity. Moreover, this design reduces the overall complexity of MDAN compared to the element-wise attention mechanism. 

\section{On-Line Scaling for Loop Filtering}

The proposed MDAN are highly capable of capturing multi-scale features and describing high-dimensional mapping. However, the off-line trained model aims to learn the average relationships between the inputs and the targets on the whole dataset and hence lacks the adaptability to some extent during inference. Moreover, there are some mismatches between the training conditions and the inference conditions in the view of video filtering. The model is usually trained based on independent frames. Nevertheless, the output frame of model in loop filtering may also serve as the reference frame for the successive to-be-coded frames.

\begin{figure}[t]
\vskip 0.2in
\begin{center}
\centerline{\includegraphics[width=7.5cm]{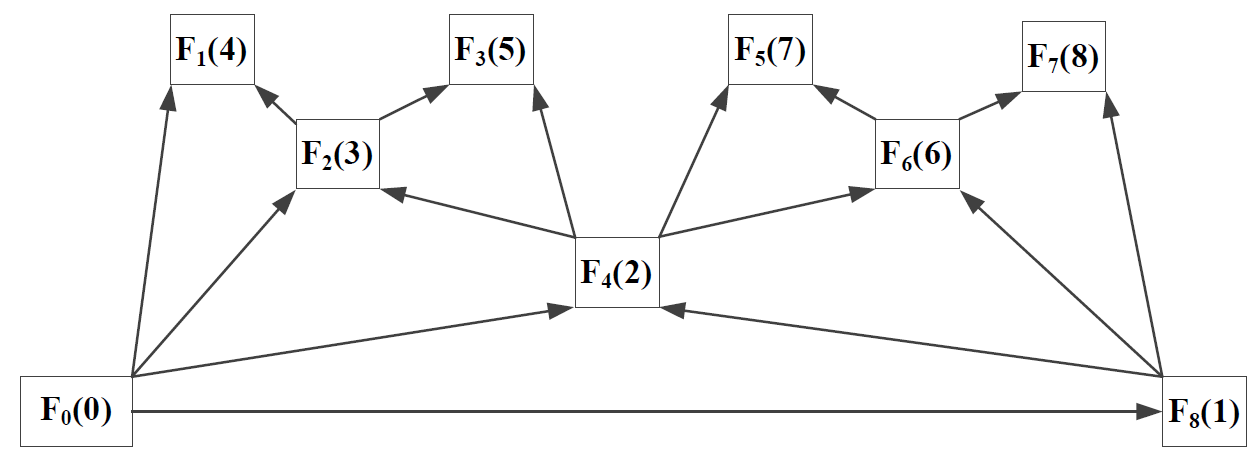}}
\caption{Illustration of the typical hierarchical-B reference structure in video coding.}
\label{fig-add}
\end{center}
\vskip -0.2in
\end{figure}

Taking the group of nine pictures as an example, the typical hierarchical-B reference structure in video compression is illustrated in Figure \ref{fig-add}, where $F_{i}(j)$ denotes the $i-th$ frame in the view of the temporal order and the $j$ represents its actual compression order. In this example, for the temporal nine frames $F_0, F_1, ... , F_8$, the compression order should be $F_0, F_8, F_4, F_2, F_1, F_3, F_6, F_7, F_8$. Therefore, for a specific frame, its reference frame has been processed by the pre-trained model, which may propagate to the corresponding reconstructed frame. If we further filter the reconstructed frame by the identical model, the behaviour differs from the mapping learned during training.  

In the loop filtering, the reconstructed frame is fed into the neural network and then the filtered reconstructed frame is obtained. Rather than using the output frame of the neural network directly, we propose to add a scaling process for the output frame, as:
\begin{equation} \label{eq:7}
P_{filter} = factor * P_{NN},
\end{equation}
where $P_{NN}$ and $P_{filter}$ represent the output frame of the neural network and the targeted filtered frame, respectively. The scaling process aims to obtain the final filtered frame owning less distortion with the original signal. Therefore, it is better to apply scaling on the difference signal between the output frame of the neural network and its corresponding original frame, and then add the output frame of the neural network to obtain the final result, as: 
\begin{equation} \label{eq:8}
\alpha * (P_{NN} - P_{rec}) + P_{rec} \rightarrow P_{org},
\end{equation}
where $P_{rec}$ and $P_{org}$ represent the reconstructed frame fed into the neural network and the original frame, respectively. 

The scaling factor$\alpha$ can be derived as an optimization problem, as
\begin{equation} \label{eq:9}
\alpha * (P_{NN} - P_{rec}) \rightarrow P_{org} - P_{rec}.
\end{equation}
According to the linear regression, the best factor $\alpha$ should be
\begin{equation} \label{eq:10}
selfMulti = \sum_{i\epsilon\Omega}(P_{NN,i}-P_{rec,i})*(P_{NN,i}-P_{rec,i}),
\end{equation}
\begin{equation} \label{eq:11}
crossMulti = \sum_{i\epsilon\Omega}(P_{NN,i}-P_{rec,i})*(P_{org,i}-P_{rec,i}),
\end{equation}
\begin{equation} \label{eq:12}
sumOrgResi = \sum_{i\epsilon\Omega}(P_{org,i}-P_{rec,i}),
\end{equation}
\begin{equation} \label{eq:13}
sumNNResi = \sum_{i\epsilon\Omega}(P_{NN,i}-P_{rec,i}),
\end{equation}
\begin{equation} \label{eq:14}
\alpha = \frac{n* crossMulti - sumOrgResi*sumNNResi}{n*selfMulti - sumNNResi*sumNNResi},
\end{equation}
where $i\epsilon\Omega$ denotes every pixel in the frame region $\Omega$ and $n$ is the total number of pixels. The derived scaling factor will be coded into the bitstream and hence the decoder can conduct the scaling process identically.  
 
\section{Experiments}

In this section, we implement the on-line scaling based MDAN as loop filtering to improve the compression efficiency beyond the latest VVC standard. Firstly, the training and inference details are given. Then, the performance of the proposed algorithm are presented and analyzed, including the objective evaluation and subjective quality. Finally, we compare the proposed MDAN with state-of-the-art algorithms to further demonstrate the advantages of our method.

\subsection{Implementation}

The number of MDSA block is a hyper-parameter and set to 8 in our experiments. The proposed MDAN is incorporated into VTM-10 \cite{VTM10}, which is the reference software of the VVC standard. 100 videos with resolution $960 \times 544$ and 100 videos with resolution $1920 \times 1088$ are randomly selected from the BVI-DVC dataset \cite{ma2020bvi} as the raw training data. The video format is YUV420, which is commonly used in mainstream video applications. In YUV420, Y channel is the luma component, and U and V channels are the chroma components. The width and height of Y channel are twice those of the U and V channels.  

\begin{figure*}[t]
\vskip 0.2in
\begin{center}
  \subfigure[Left: VTM10, 37.6517 dB; right: Proposed, 38.0223 dB.]
  {\includegraphics[width=4.1cm, height = 2.7cm]{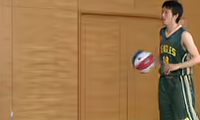}
  \includegraphics[width=4.1cm, height = 2.7cm]{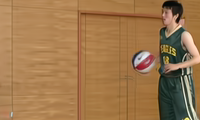} }
  \subfigure[Left: VTM10, 36.4576 dB; right: Proposed, 36.9522 dB.]
  {\includegraphics[width=4.1cm, height = 2.7cm]{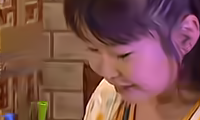}
  \includegraphics[width=4.1cm, height = 2.7cm]{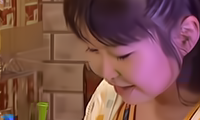} }
  \subfigure[Left: VTM10, 32.7639 dB; right: Proposed, 33.3483 dB.]
  {\includegraphics[width=4.1cm, height = 2.7cm]{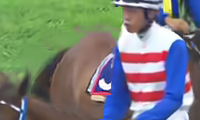}
  \includegraphics[width=4.1cm, height = 2.7cm]{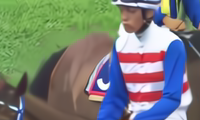} }
  \subfigure[Left: VTM10, 32.8397 dB; right: Proposed, 33.4650 dB.]
  {\includegraphics[width=4.1cm, height = 2.7cm]{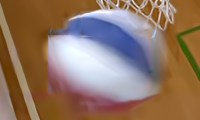}
  \includegraphics[width=4.1cm, height = 2.7cm]{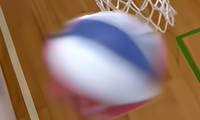} }
  \subfigure[Left: VTM10, 35.3360 dB; right: Proposed, 35.9541 dB.]
  {\includegraphics[width=4.1cm, height = 2.7cm]{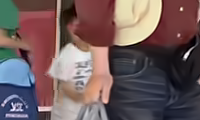}
  \includegraphics[width=4.1cm, height = 2.7cm]{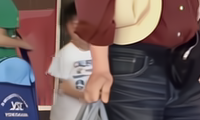} }
  \subfigure[Left: VTM10, 33.6547 dB; right: Proposed, 34.0581 dB.]
  {\includegraphics[width=4.1cm, height = 2.7cm]{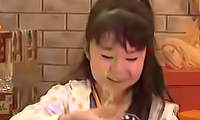}
  \includegraphics[width=4.1cm, height = 2.7cm]{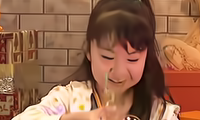} }
\caption{Visual quality comparisons between VTM-10 and the proposed algorithms. (a) \emph{BasketballPass}, (b) \emph{BlowingBubbles}, (c) \emph{RaceHorses}, (d) \emph{BasketballDrill}, (e) \emph{BQMall}, (f) \emph{PartyScene}.}
\label{fig4}
\end{center}
\vskip -0.2in
\end{figure*}

The raw training videos are compressed by VTM-10 at different bitrates, controlled by quantization parameter (QP) (22, 27, 32, 37, 42). The reconstructed frames and the corresponding original frames are collected as the input and target to be learned. Because using one fixed off-line trained model cannot sufficiently describe the mapping relationships between the reconstructed frames of different quality under various bitrates and the same original frame, we split the reconstructed frames into five subsets according to the QP, and one model is trained for each subset. During inference, the model is selected according to the QP of each reconstructed frame. The reconstructed frame is input to MDAN and the output frame will be further processed by the scaling factor. In this manner, the compression distortion and artifacts of the reconstructed frame can be reduced, which also provides better reference signal for the compression of subsequent frames.

\subsection{Performance of the on-line scaling based MDAN for Loop Filtering}

The experiments are conducted following the JVET common test conditions (CTC) \cite{CTC} under random access configuration for QPs (22, 27, 32, 37). The test sequences contains various content and different resolutions including $3840 \times 2160$, $1920 \times 1080$, $832 \times 480$ and $416 \times 240$. The compression performance is evaluated by Bjontegaard delta rate (BD-rate) and Bjontegaard delta PSNR (BD-PSNR) \cite{Bjontegaard}. BD-rate measures the average bit saving at the same PSNR between the anchor and the proposed method. Similarly, BD-PSNR measures the average PSNR difference between two methods at the same bit rate.  

The performance of the proposed MDAN is shown in Table \ref{table1}, from which it is observed that on average 6.08\%, 15.50\% and 15.61\% bitrate savings can be achieved for Y, U and V components, respectively. If we adopt the factors 4:1:1 as weights of the bitrate savings for the Y, U and V components, then an overall 9.24\% bitrate saving can be obtained. Observing the results on each test sequence, it is noticed that promising performance is obtained on all sequences. The proposed MDAN also performs well on the sequences with resolutions of $3840 \times 2160$ and $416 \times 240$, which are far from the resolutions used in the training dataset. This observation proves the superiority and robustness of the proposed MDAN, benefiting from the multi-density fusion method and the combination of spatial and channel attention mechanism.

Furthermore, the on-line scaling method is incorporated with MDAN and the performance in terms of BD-rate are presented in the right column in Table \ref{table1}. It is found that additional 0.49\%, 1.67\% and 2.02\% coding gain are achieved for Y, U and V components respectively and the overall bitrate saving can reach up to 10.18\%. When we assess the compression performance in terms of PSNR improvement, it is observed that the proposed algorithms bring 0.255 dB, 0.577 dB and 0.598 dB improvements for the Y, U and V components, and the overall quality improvement is 0.366 dB on average after weighting.  

\begin{table*}[tbp]
\caption{Comparisons with the state-of-the-art methods.}
\label{table2}
\begin{tabular}{cccccccccc}
\toprule
\textbf{Sequence} & \textbf{Zhang'19} & \textbf{Zamir'20} & \textbf{Pan'20} & \textbf{JVET-T0079} & \textbf{JVET-T0088} & \textbf{JVET-T0128} & \textbf{Proposed}  \\ 
\midrule
  \emph{Tango2}          & -8.39\% & -6.17\% & -7.17\% & -7.03\% & -6.15\% & -8.00\%  & -11.24\%  \\
  \emph{FoodMarket4}     & -10.34\% & -3.96\% & -3.97\% & -3.77\% & -4.77\% & -8.45\% & -9.45\%  \\
  \emph{Campfire}        & -7.22\% & -6.57\% & -3.99\% & -5.27\% & -1.70\% & -6.99\%  & -9.90\%  \\
  \emph{CatRobot}        & -9.27\% & -7.77\% & -3.08\% & -8.18\% & -8.90\% & -7.79\%  & -10.60\%  \\
  \emph{DaylightRoad2}   & -6.51\% & -7.84\% & -3.29\% & -8.86\% & -13.55\% & -8.45\% & -11.84\%  \\
  \emph{ParkRunning3}    & -4.12\% & -2.46\% & -3.68\% & -3.44\% & -2.41\% & -0.84\%  & -5.67\%  \\
  \emph{MarketPlace}     & -7.11\% & -7.40\% & -4.85\% & -6.85\% & -7.90\% & -8.56\%  & -9.45\%  \\
  \emph{RitualDance}     & -3.62\% & -8.01\% & -8.65\% & -4.53\% & -5.04\% & -9.73\%  & -11.72\%  \\
  \emph{Cactus}          & -6.24\% & -5.58\% & -5.37\% & -6.02\% & -6.90\% & -8.14\%  & -8.82\%  \\
  \emph{BasketballDrive} & -4.25\% & -5.82\% & -7.28\% & -5.98\% & -5.34\% & -8.21\%  & -10.30\%  \\
  \emph{BQTerrace}       & -3.07\% & -6.78\% & -4.34\% & -8.15\% & -9.90\% & -5.73\%  & -9.07\%  \\
  \emph{BasketballDrill} & -4.20\% & -7.87\% & -5.54\% & -5.62\% & -6.91\% & -4.16\%  & -10.37\%  \\
  \emph{BQMall}          & -5.32\% & -8.92\% & -8.99\% & -7.10\% & -8.38\% & -5.83\%  & -11.20\%  \\
  \emph{PartyScene}      & -8.12\% & -6.89\% & -2.52\% & -4.39\% & -5.55\% & -2.87\%  & -9.96\%  \\
  \emph{RaceHorses}      & -8.40\% & -7.35\% & -8.53\% & -5.85\% & -4.39\% & -5.43\%  & -9.22\%  \\
  \emph{BasketballPass}  & -7.10\% & -9.16\% & -6.22\% & -5.89\% & -4.79\% & -4.56\%  & -10.37\% \\
  \emph{BQSquare}        & -5.49\% & -10.27\% & -5.00\% & -7.92\% & -11.13\% & -2.38\%& -12.77\%  \\
  \emph{BlowingBubbles}  & -3.16\% & -7.69\% & -5.13\% & -4.46\% & -6.47\% & -3.71\%  & -9.80\%  \\
  \emph{RaceHorses}      & -4.64\% & -10.61\% & -3.28\% & -6.57\% & -5.60\% & -4.46\% & -11.68\%  \\
\midrule
\textbf{Average} & \textbf{-6.14\%} & \textbf{-7.22\%} & \textbf{-5.31\%} & \textbf{-6.29\%} & \textbf{-6.60\%} & \textbf{-6.02\%} & \textbf{-10.18\%}  \\ 
\bottomrule
\end{tabular}
\end{table*}

The visual quality comparisons are presented in Figure \ref{fig4}, where the same frames are compressed by the Anchor (VTM-10) and the proposed algorithms at similar bitrates, respectively. Regarding the visual quality, from the decoded frames we can obviously observe that the proposed method can reduce the compression distortion and preserve more textural details, e.g., the wooden wall (Figure \ref{fig4}(a)), the grassland (Figure \ref{fig4}(c)), the t-shirt (Figure \ref{fig4}(e)), etc. Moreover, the proposed algorithms eliminates the compression artifacts to a large extent, such as the suppression of ringing (Figure \ref{fig4}(b)) artifact, blocking artifact (Figure \ref{fig4}(d)), and color shift (Figure \ref{fig4}(f)). 

\subsection{Performance Comparisons with the State-of-the-Art Methods}

We further compare the proposed algorithms with other state-of-the-art methods. The recently advanced works are reproduced for loop filtering under the same training and test conditions, including the residual non-local attention network \cite{zhang2019residual}, multi-scale residual network \cite{zamir2020learning} and enhanced residual network \cite{pan2020efficient}. Moreover, the prominent proposals of neural network based loop filtering during the JVET activities are also included, e.g., JVET-T0079 \cite{JVET-T0079}, JVET-T0088 \cite{JVET-T0088} and JVET-T0128 \cite{JVET-T0128}. The BD-rates for YUV with 4:1:1 weighting of all methods are shown in Table \ref{table2}. It is observed that the proposed algorithms achieves significant improvement of compression efficiency compared to the state-of-the-art methods. On one hand, the outstanding performance of MDAN owes to the capability to extract rich features from multiple densities, the refinement of feature map by attention branch, and the novel fusion mechanism. On the other hand, the on-line scaling method enhances the signal adaptability of the proposed scheme and further reduces the compression distortion by utilizing the original signal.  

\section{Conclusion}

In this paper, we introduce a novel multi-density attention network which adopts multi-density single-attention structure. To achieve precise presentation of signal and the larger receptive field simultaneously, we first propose to apply multi-density branches to extract rich features from different sampling densities. Then, a single-path spatial attention branch is introduced to learn the spatial sample correlations and refine the feature maps. Furthermore, an advanced fusion procedure based on channel-mutual attention is proposed, which can realize data exchange and fusion more adaptively. Finally, the on-line scaling process is incorporated to optimize the output results of MDAN and further reduce the compression distortion and artifacts. Our experiments demonstrate the performance and advantages of the proposed algorithms, which can achieve 10.18\% bitrate saving for video compression over the latest video coding standard VVC and outperforms other state-of-the-art methods. 

This work is, to the best of our knowledge, the first to propose a multi-density model with separate spatial- and channel- attention. It is also an interesting study to combine the off-line pre-trained model and the on-line scaling learning into video compression and achieve better ability of signal processing. We believe that the principles of our algorithms (multi-density extraction, single attention branch to generate multi mask maps, channel-mutual data fusion, on-line scaling process) can be generally applied to other works as well. Interesting future works include the investigation of the number and location of MDSA blocks and light weight implementation of multi-density learning to reduce complexity.

\bibliographystyle{ACM-Reference-Format}
\bibliography{sample-base}

\end{document}